\documentclass[fleqn,usenatbib]{mnras}

\usepackage{newtxtext,newtxmath}
\usepackage[T1]{fontenc}

\DeclareRobustCommand{\VAN}[3]{#2}
\let\VANthebibliography\thebibliography
\def\thebibliography{\DeclareRobustCommand{\VAN}[3]{##3}\VANthebibliography}

\usepackage{graphicx}	% Including figure files
\usepackage{amsmath}	% Advanced maths commands
\newcommand{\hi}{\mbox{H\,{\sc i}}}

\def\ha{H\,$\alpha$}

\def\oiii{[O\,{\sc iii}]}

\def\neiii{[Ne\,{\sc iii}]}
\def\nev{[Ne\,{\sc v}]}

\def\kms{km\,s$^{-1}$}
\def\cms{cm$^{-2}$}

\def\zabs{$z_{\rm abs}$}
\def\zem{$z_{\rm em}$}
\def\nhi{$N$($\hi$)}
\def\ts{T$_{\rm s}$}
\def\fc{f$_{\rm c}$}
\def\mjb{mJy~beam$^{-1}$}
\def\taudv{$\int\tau dv$}
\def\taup{$\tau_{\rm p}$}
\def\v90{$v_{\rm 90}$}

\title[\hi\ in double-peaked AGN]{\hi\ 21-cm absorption in radio-loud AGN with double-peaked \oiii\ emission}

\author[Dutta et al.]{
Rajeshwari Dutta$^{1,2}$\thanks{rajeshwari.dutta@unimib.it} 
and Raghunathan Srianand$^{3}$
\\
$^{1}$Dipartimento di Fisica G. Occhialini, Universit\`a degli Studi di Milano Bicocca, Piazza della Scienza 3, 20126 Milano, Italy \\
$^{2}$INAF - Osservatorio Astronomico di Brera, via Bianchi 46, 23087 Merate (LC), Italy \\
$^{3}$ IUCAA, Postbag 4, Ganeshkhind, Pune 411007, India \\
}

\date{Accepted XXX. Received YYY; in original form ZZZ}

\pubyear{2022}

\begin{document}
\label{firstpage}
\pagerange{\pageref{firstpage}--\pageref{lastpage}}
\maketitle

\begin{abstract}

Different physical processes in galaxy evolution, such as galaxy mergers that lead to coalescence of dual Active Galactic Nuclei (AGN) and outflows emanating from the narrow line region, can leave their imprint on the optical spectra of AGN in the form of double-peaked narrow emission lines. To investigate the neutral gas in the centres of such AGN, we have conducted a pilot survey of \hi\ 21-cm absorption, using the upgraded Giant Metrewave Radio Telescope (uGMRT), in radio-loud AGN whose optical spectra show double-peaked \oiii\ emission lines at $z\le0.4$ (median $z\approx0.14$). Among the eight sources for which we could obtain clean spectra, we detect \hi\ 21-cm absorption in three sources (detection rate of $38^{+36}_{-20}$\%) and find tentative indication of absorption in two other sources. The detection rate of \hi\ 21-cm absorption is tentatively higher for the systems that show signatures of interaction or tidal disturbance ($\gtrsim50$\%) in the ground-based optical images than that for the systems that appear single and undisturbed ($\approx25$\%). This is consistent with the high incidence of \hi\ 21-cm absorption observed in $z\le0.2$ galaxy mergers. Higher spatial resolution spectroscopy is required to confirm the origin of the \hi\ absorbing gas, i.e. either gas infalling onto the radio-loud AGN, outflowing gas ejected by the AGN, or gas in rotation on the galactic-scale or circumnuclear discs.

\end{abstract}


%%%%%%%%%%%%%%%%%%%%%%%%%%%%%%%%%%%%%%%%%%%%%%%%%%

%%%%%%%%%%%%%%%%% BODY OF PAPER %%%%%%%%%%%%%%%%%%

\section{Introduction}
\label{sec_intro}

Mergers play a vital role in the evolution of galaxies. In addition to significantly impacting the physical conditions of gas  and the star formation rate within galaxies, mergers can initiate Active Galactic Nuclei (AGN) in the centres of galaxies \citep{springel2005,hopkins2006}. Both galaxy mergers and interactions of the AGN with the host galaxies can leave their imprint on the emission line spectra of galaxies. During a merger of two galaxies, each with a central super-massive black hole (SMBH) and sufficient nuclear gas inflow, both of the SMBHs can trigger AGN \citep{begelman1980,comerford2009a,blecha2013}. Consequently, the orbital motion of the dual AGN, as they spiral towards each other, leads to two spatially and kinematically separated emission lines or a double-peaked emission line profile in the spatially-integrated spectrum of the galaxy. Dual AGN represent the stage in a merger before coalescence and are important for studying the connection between galaxy mergers and AGN activity. Moreover, they could be sources of gravitational wave emission at later stages of coalescence \citep{flanagan1998,klein2016}. 

On the other hand, energy from a single AGN can drive bi-conical outflows from the narrow line region (NLR; $\sim$100\,pc), producing similar double-peaked emission line profiles \citep[e.g.][]{veilleux2001,fischer2011}. Possible mechanisms responsible for driving the outflows include radiation pressure acting on gas and dust, and hot winds or radio jets entraining the NLR clouds. This AGN feedback plays an important role in quenching the star formation, that is required to reproduce the observed galaxy properties \citep{croton2006}, and in establishing the observed correlation between the SMBH mass and galaxy mass or velocity dispersion \citep{ferrarese2000}. Other possible physical origins of the double-peaked emission lines include complex NLR kinematics and rotating star-forming discs \citep[e.g.][]{shen2011,nevin2016}. Therefore, AGN which exhibit double-peaked emission lines are valuable tools for investigating several aspects of AGN and galaxy co-evolution, including AGN triggering and feedback.

The double-peaked emission lines, i.e. those that are best fit with two components instead of one, are usually more pronounced in forbidden transition lines of high ionization energy such as \oiii\ $\lambda$5007, \neiii\ $\lambda$3869 and \nev\ $\lambda$3426. Double-peaked \ha\ emission lines have also been identified in samples of radio-loud AGN and have been proposed to arise due to accretion disk emission \citep[e.g.][]{eracleous2003}. In recent years thanks to spectroscopic surveys like the Sloan Digital Sky Survey \citep[SDSS;][]{york2000}, several hundred double-peaked narrow emission lines have been identified and they account for $\approx$1\% of the AGN population \citep[e.g.][]{wang2009,liu2010,smith2010,barrows2013}. In particular, searches for double-peaked AGN at $z\le0.8$ have utilized the strong emission line of \oiii$\lambda$5007. Since the projected angular size of the SDSS fibre is 3 arcsec (2 arcsec in the case of SDSS-BOSS observations), higher spatial-resolution observations are usually required to determine the origin of the double-peaked lines. Follow-up observations, such as high spatial-resolution optical/near-infrared (NIR) imaging and spectroscopy, X-ray and radio observations, have confirmed the presence of dual AGN or outflows in some cases \citep[e.g.][]{comerford2011,comerford2018,liu2018a,liu2018b,rubinur2019,nandi2021,kharb2021}. 

The circumnuclear neutral gas in radio-loud AGN can be probed using \hi\ 21-cm absorption \citep[see for a review][]{morganti2018}. Associated \hi\ absorption has been used to study: (i) feedback from AGN in the form of neutral gas outflows (which could be induced by radio jets), (ii) cold interstellar medium in the central regions of AGN, and (iii) fueling of AGN (which could be through neutral gas funneled into the central regions due to mergers). If detected, \hi\ 21-cm absorption can be useful to probe the circumnuclear neutral gas at parsec-scales through Very Long Baseline Array (VLBA) spectroscopy \citep[e.g.][]{srianand2015}. Hundreds of radio-loud AGN have been searched for \hi\ 21-cm absorption till date, with typical detection rates of $\approx20-30$\% at $z\lesssim0.2$ \citep[e.g.][]{vangorkom1989,morganti2001,curran2006,gupta2006,allison2012,gereb2015,maccagni2017,chandola2020}, and lower detection rates ($\lesssim10$\%) at higher redshifts \citep[e.g.][]{curran2008,aditya2016,aditya2018,grasha2019,Gupta2021,murthy2022,su2022}. Recently, it has been shown that the incidence of \hi\ 21-cm absorption is elevated in radio-loud AGN that are part of merging systems at $z\le0.2$, with a detection rate of $\approx$84\% \citep{dutta2018,dutta2019}. The significant increase in the incidence and column density of neutral hydrogen gas in such AGN indicate that the merger process is likely to be successful in funnelling large quantities of neutral gas to the central regions of these galaxies and consequently could be playing a role in activating the radio-loud AGN.

In the observations of double-peaked AGN mentioned above, the kinematics of the ionized gas have been well-studied. To summarize, these studies have proposed different physical mechanisms for the origin of the double-peaked emission, including dual AGN in a merging system \citep[e.g.][]{comerford2009b,Shangguan2016,liu2018b}, outflows \citep[induced by radio jets or other mechanisms; e.g.][]{rosario2010b,fischer2011,greene2012}, NLR kinematics \citep[e.g.][]{shen2011}, and rotating rings/discs of gas \citep[e.g.][]{smith2012}. However, what have not yet been investigated are the properties and kinematics of the cold neutral gas in these special type of AGN. Hence, searching for \hi\ 21-cm absorption in a sample of double-peaked AGN would be instrumental in characterizing the neutral gas properties in different stages of AGN evolution, like triggering of AGN through mergers and feedback from AGN in the form of outflows.

We present here a pilot survey of \hi\ 21-cm absorption in double-peaked radio-loud AGN using the upgraded Giant Metrewave Radio Telescope \citep[uGMRT;][]{gupta2017}. The details of the sample and the radio observations are presented in Section~\ref{sec_sample_obs}. The results from this sample and the \hi\ 21-cm absorption associated with individual sources are presented and discussed in Section~\ref{sec_results}. The results are discussed and summarized in Section~\ref{sec_summary}. We adopt a Planck 15 cosmology with $H_{\rm 0}$ = 67.7\,\kms\,Mpc$^{-1}$ and $\Omega_{\rm M}$ = 0.307 throughout this paper \citep{planck2016}.

\section{Sample and Observations}
\label{sec_sample_obs}
\subsection{Sample Selection and Properties}
\label{sec_sample}

To obtain the sample of double-peaked radio-loud AGN, we cross-matched compilations of AGN showing double-peaked \oiii\ $\lambda$5007 emission line in SDSS spectra \citep{wang2009,liu2010} with radio sources from the Faint Images of the Radio Sky at Twenty-Centimeters \citep[FIRST;][]{white1997} and NRAO VLA Sky Survey \citep[NVSS;][]{condon1998}. We restricted to AGN whose redshifted \hi\ 21-cm line is covered by Band-5 of uGMRT, i.e. $z\le0.4$, and to radio sources with flux density at 1.4\,GHz ($S_{\rm 1.4~GHz}$) greater than 30\,mJy for sensitive \hi\ 21-cm absorption search. After visually inspecting the radio images and optical spectra, we obtained a sample of ten double-peaked AGN at $z\le0.4$, having a median redshift of $\approx0.14$. The velocity separation between the two Gaussian components of the double-peaked \oiii\ line ranges between $\approx$280\,\kms\ and $\approx$740\,\kms\ with a median of $\approx$400\,\kms. The full-width at half-maximum of the individual components ranges between $\approx$60\,\kms\ and $\approx$700\,\kms\ with a median of $\approx$250\,\kms.

Out of the ten double-peaked AGN, five (J0738+3156, J0842+0547, J1203+1319, J1243-0058 and J1517+3353) appear as single, undisturbed systems in the available optical images [Dark Energy Camera Legacy Survey \citep[DECaLS;][]{dey2019}, PanSTARRS-1 \citep{chambers2016} and SDSS]. The remaining AGN are either part of galaxy pairs and interacting systems or show disturbed morphology. J0009-0036 has a companion galaxy at a projected separation of $\approx$40\,kpc and a velocity separation of $\approx$460\,\kms. J1526+4140 is undergoing merger with another galaxy at $\approx$5\,kpc and $\approx$150\,\kms. The optical continuum image of J0912+5320 shows two peaks and a tidal tail, while that of J1352+6541 shows signatures of a possible tidal disturbance. J1356+1026 is a post-merger system that has been observed as part of our \hi\ 21-cm absorption survey in low-$z$ galaxy mergers \citep{dutta2018}. Long-slit spectroscopy and NIR imaging of this system show double nuclei separated by 2.9\,kpc \citep{greene2011,shen2011}. Moreover, there have been observations of extended \oiii\ emission in this system, suggesting large-scale outflows powered by feedback from the obscured, radio-quiet quasar \citep{greene2012}.

Three of the AGN that appear as single systems (J0738+3156, J0842+0547 and J1203+1319) have been observed with the Keck Laser Guide Star Adaptive Optics system and the NIR camera NIRC2 \citep{mcgurk2015}. Based on these imaging observations, J1203+1319 has a double spatial structure on scales of $<3$ arcsec, indicating it is a candidate dual AGN. J0738+3156 and J0842+0547, on the other hand, show single spatial structure with no companions within 3 arcsec. Furthermore, long-slit optical spectra have been obtained for four AGN (J0738+3156, J0842+0547, J1243-0058 and J1517+3353) by \citet{comerford2018}. Based on these spectra, J0842+0547 and J1243-0058 are classified as outflows, whereas J1517+3353 is classified as an outflow composite, i.e. an outflow that includes many gas clouds at different velocities. On the other hand, J0738+3156 has a disturbed rotation-dominated profile.

The 1.4\,GHz radio continuum emission of all the AGN appear compact (deconvolved sizes $\lesssim$2 arcsec) in the FIRST images, with the peak flux density accounting for $\gtrsim$90\% of the total flux density. Based on  high-resolution ($\approx$0.2 arcsec) multi-band (8-12 GHz) VLA observations \citep{muller2015}, J0009-0036 shows a two-sided radio jet suggesting that a radio jet-driven outflow could produce the double-peaked emission. In addition, six of the AGN have been imaged with VLBA at milliarcsecond (mas; $\lesssim$10\,pc) resolution \citep{liu2018a}. Three of them (J0009-0036, J0738+3156 and J1356+1026) remain undetected in the VLBA 8.4 GHz images. J0912+5320 shows three components in the VLBA image, an unresolved central core and northern and southern components extending over $\approx$25 mas, indicative of a jet structure. J1243-0058 shows a compact, unresolved component in the VLBA image, and there are no other radio sources stronger than 1\,mJy within 1 arcsec. J1352+6541 shows a faint component at 2.4 mas from the primary component.

\subsection{Observations and Data Reduction}
\label{sec_obs}

\begin{table*}
\caption{Results from the \hi\ 21-cm observations of the radio-loud AGN showing double-peaked \oiii\ emission. 
Column 1: name used in this work. 
Column 2: coordinates of the source.
Column 3: average redshift of the double-peaked \oiii\ emission line.
Column 4: redshift of the peak \hi\ 21-cm absorption in case of detection.
Column 5: peak flux density of the radio source (there is an additional nearby radio source in the case of J0912+5320 and J1526+4140).
Column 6: velocity width of spectral channel.
Column 7: spectral rms per velocity channel.
Column 8: peak \hi\ 21-cm optical depth in case of detection or $3\sigma$ upper limit per velocity channel in case of non-detection.
Column 9: integrated \hi\ 21-cm optical depth in case of detection or $3\sigma$ upper limit for a velocity width of 100\,\kms\ in case of non-detection.
Column 10: \nhi\ for a spin temperature of 100\,K and unit covering factor.
Note that the results for the tentative detections (J0009-0036 and J0912+5320) are listed in italics.
}
\centering
\begin{tabular}{cccccccccc}
\hline
Name & Coordinates & \zem\ & \zabs\ & Peak 1.4\,GHz & $\delta$v & Spectral & \taup\ & \taudv\ & \nhi\                   \\
     & (J2000)     &       &        & Flux Density  & (\kms)    & rms      &        & (\kms)  & (\ts/100\,K)(1/\fc)     \\
     &             &       &        & (\mjb)        &           & (\mjb)   &        &         & ($\times10^{20}$\,\cms) \\
 (1) & (2)         & (3)   & (4)    & (5)           & (6)       & (7)      & (8)    & (9)     & (10)                    \\     
\hline
J0009-0036 & 00 09 11.58 -00 36 54.70 & 0.07333 & {\it 0.07324} & 39 & 2.8 & 1.4 & {\it 0.12$\pm$0.04} & {\it 1.5$\pm$0.5} & {\it 3$\pm$1} \\  
J0738+3156 & 07 38 49.75 +31 56 11.82 & 0.29734 & \multicolumn{7}{c}{---RFI---} \\
J0842+0547 & 08 42 27.46 +05 47 16.40 & 0.27458 & ---     &  49 & 3.3 &  1.3 & $\le$0.09     & $\le$1.6     & $\le$3    \\
J0912+5320 & 09 12 01.68 +53 20 36.60 & 0.10173 & {\it 0.10183} & 98 & 1.4 & 2.7 & {\it 0.14$\pm$0.03} & {\it 1.8$\pm$0.3} & {\it 3$\pm$1} \\
           &                          &         & ---     &  28 & 1.4 &  2.9 & $\le$0.33     & $\le$4.2     & $\le$8    \\
J1203+1319 & 12 03 20.70 +13 19 31.00 & 0.05837 & ---     &  98 & 1.4 &  3.3 & $\le$0.09     & $\le$1.3     & $\le$2    \\
J1243-0058 & 12 43 58.36 -00 58 45.40 & 0.40893 & ---     &  48 & 3.7 &  1.1 & $\le$0.06     & $\le$1.5     & $\le$3    \\
J1352+6541 & 13 52 51.22 +65 41 13.10 & 0.20681 & \multicolumn{7}{c}{---RFI---} \\
J1356+1026 & 13 56 46.12 +10 26 09.09 & 0.12313 & 0.12369 &  72 & 8.0 &  1.7 & 0.20$\pm$0.02 & 26.0$\pm$1.0 & 47$\pm$2  \\ 
J1517+3353 & 15 17 09.23 +33 53 24.60 & 0.13515 & 0.13563 & 100 & 1.5 &  3.5 & 0.52$\pm$0.04 & 84.8$\pm$0.9 & 155$\pm$2 \\  
J1526+4140 & 15 26 06.16 +41 40 14.39 & 0.00829 & 0.00787 &  56 & 2.6 &  1.2 & 0.32$\pm$0.02 & 17.3$\pm$0.5 & 32$\pm$1  \\
           &                          &         & ---     &  19 & 2.6 &  1.2 & $\le$0.18     & $\le$2.8     & $\le$5    \\
\hline
\end{tabular}
\label{tab_radio_abs}
\end{table*}

We observed nine of the double-peaked radio sources with Band-5 (1000-1460\,MHz) of uGMRT in July 2019 (Proposal ID: 36\_016). The observing time of each source ranged between 2 to 9 hours. Standard calibrators were observed in regular intervals during the course of the observations for the purpose of calibrating the flux density, bandpass, and gain. We obtained the data in two parallel hand correlations. We used a bandwidth of 12.5\,MHz, which gives a velocity coverage of $\approx2700-3700$\,\kms\ at the redshift of the sources in our sample. Further, we divided the bandwidth into 2048 channels, which gives a velocity resolution of $\approx1.5-3.5$\,\kms\ at the redshift of the sources. One of the double-peaked sources, J1356+1026, was observed using GMRT (Proposal ID: 33\_027), as part of our \hi\ 21-cm absorption survey in low-$z$ mergers \citep{dutta2018}, using a bandwidth of 16\,MHz split into 512 channels (velocity resolution $\approx8$\,\kms). 

The data reduction was carried out using the Astronomical Image Processing System \citep[{\sc aips};][]{greisen2003} following standard procedures \citep[see][]{dutta2016}. The spectral cubes were obtained by imaging the continuum-subtracted data using "ROBUST=0" weighting. The typical spatial resolution of the data is $\approx2-3$ arcsec. The continuum emission of the radio sources is compact at this scale, with $\approx80-100$\% of the total emission contained within a single Gaussian component. The \hi\ 21-cm absorption spectra were extracted at the location of the peak continuum flux density of the radio sources (see Fig.~\ref{fig:sdss_gmrt_spec}). In the case of J0912+5320 and J1526+4140, there is a weaker radio source nearby at 34 arcsec ($\approx$65\,kpc) and 28 arcsec ($\approx$5\,kpc) separation, respectively, towards which we also extract spectra.

\section{Results}
\label{sec_results}

\begin{figure*}
    \centering
    \includegraphics[width=1.0\textwidth]{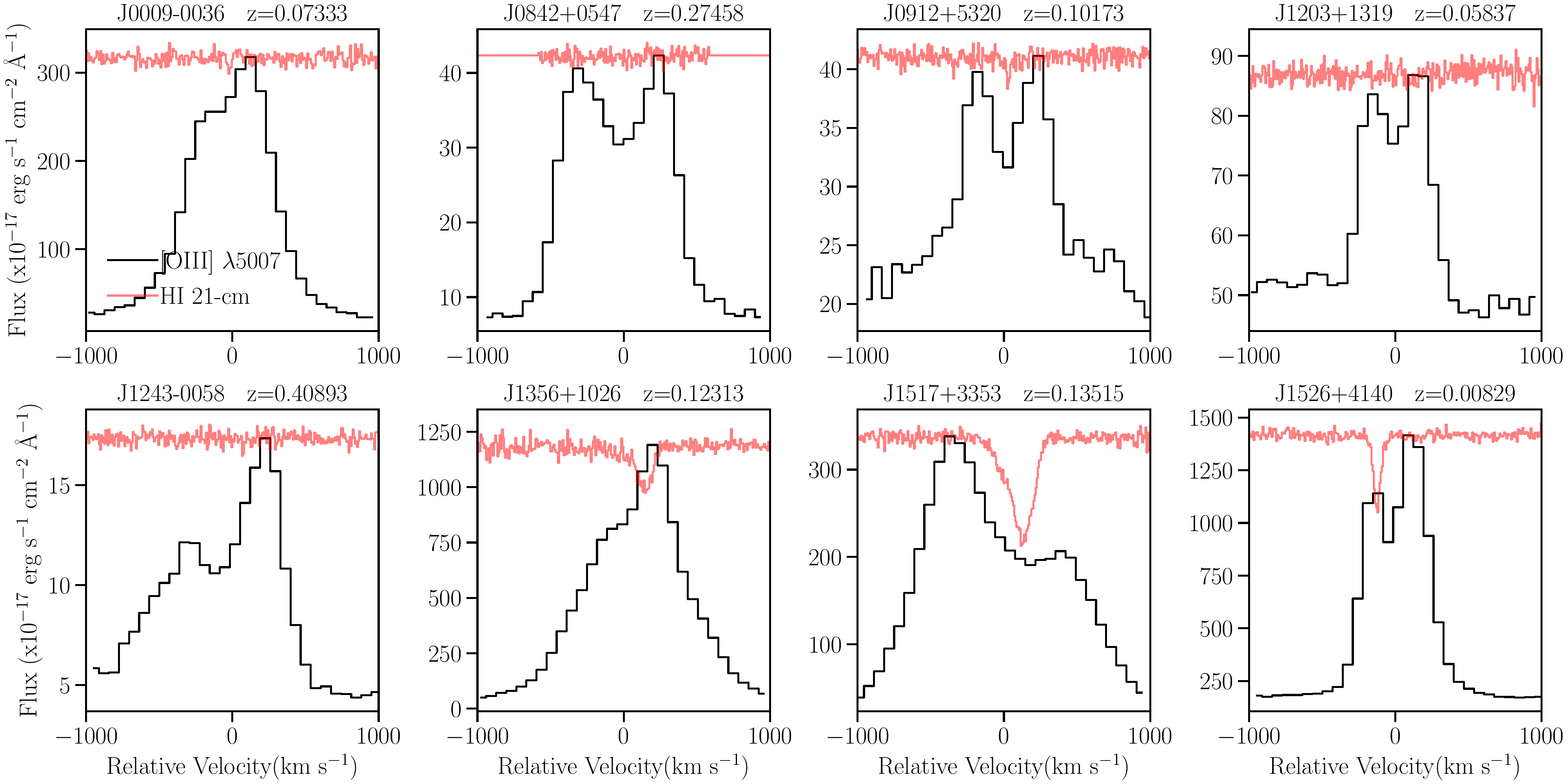}
    \caption{SDSS spectra showing the double-peaked \oiii\ emission line, of the sources for which we could obtain clean \hi\ 21-cm spectra, are plotted in black. Overplotted in arbitrary flux scale are the \hi\ 21-cm absorption spectra in pink, smoothed to $\approx8$\,\kms\ velocity bins for display purpose. The x-axis is the velocity relative to the redshift given at the top of each panel.}
    \label{fig:sdss_gmrt_spec}
\end{figure*}

\begin{figure*}
    \centering
    \includegraphics[width=0.32\textwidth]{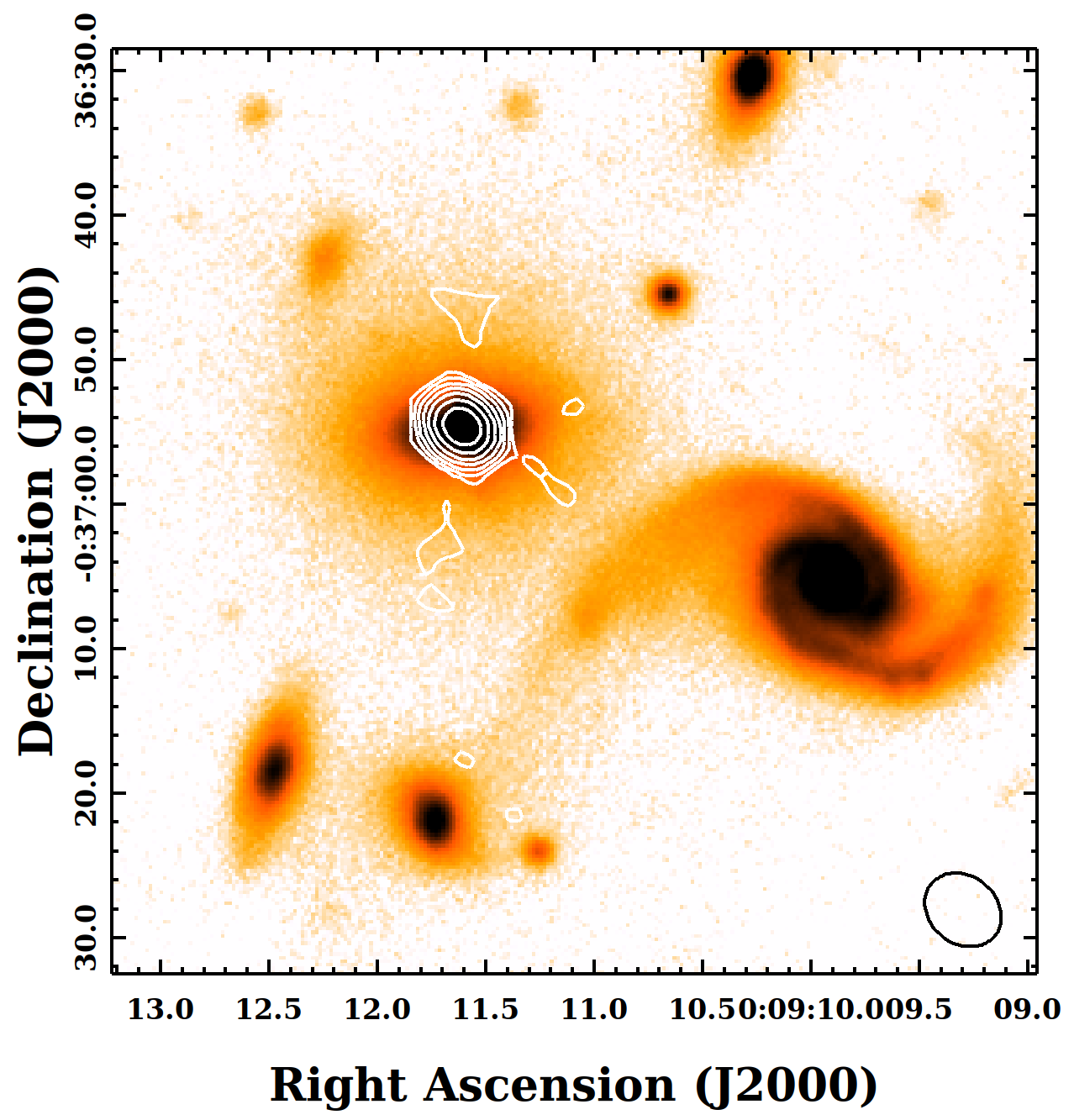}
    \includegraphics[width=0.48\textwidth]{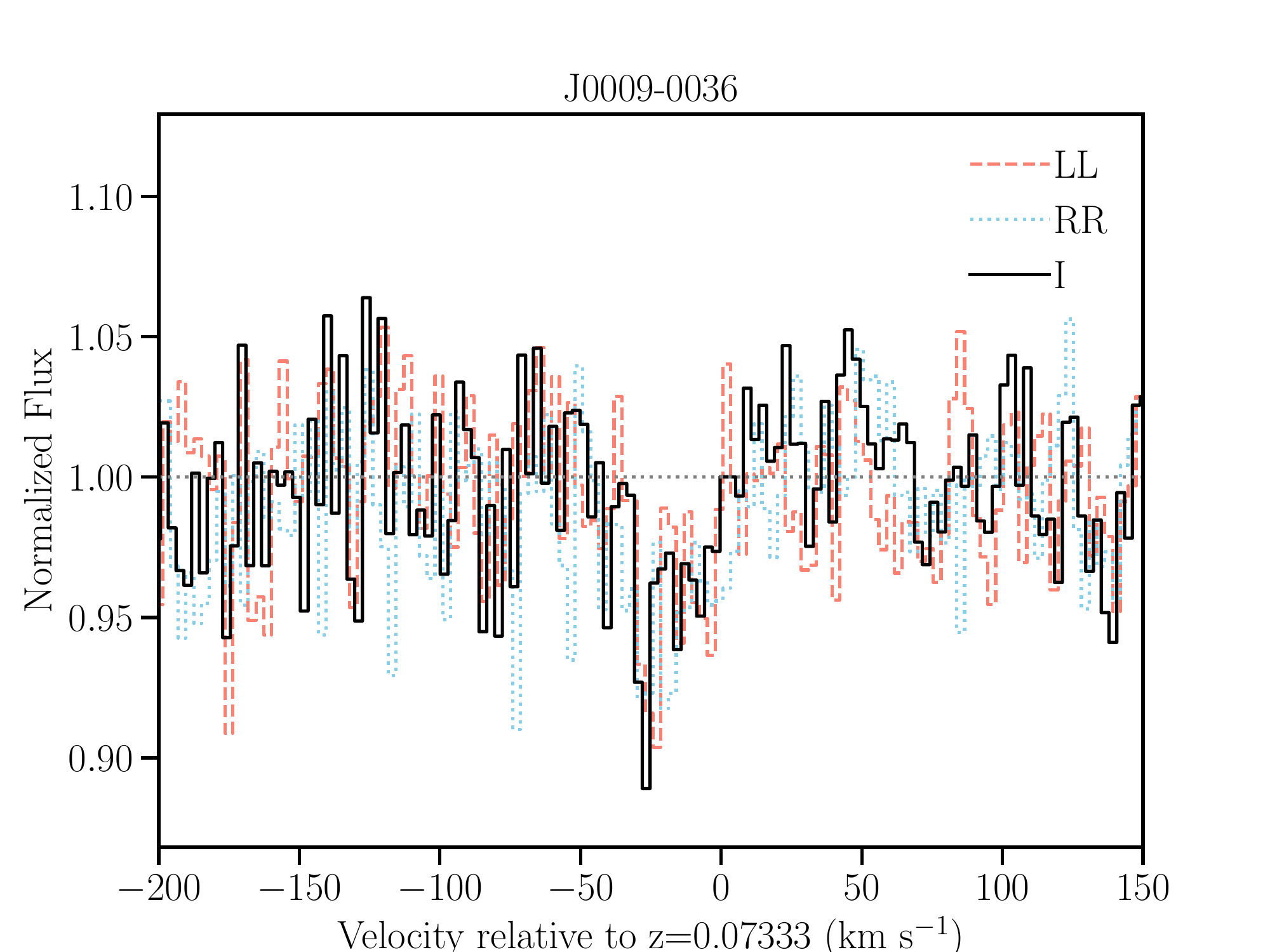}
    \includegraphics[width=0.32\textwidth]{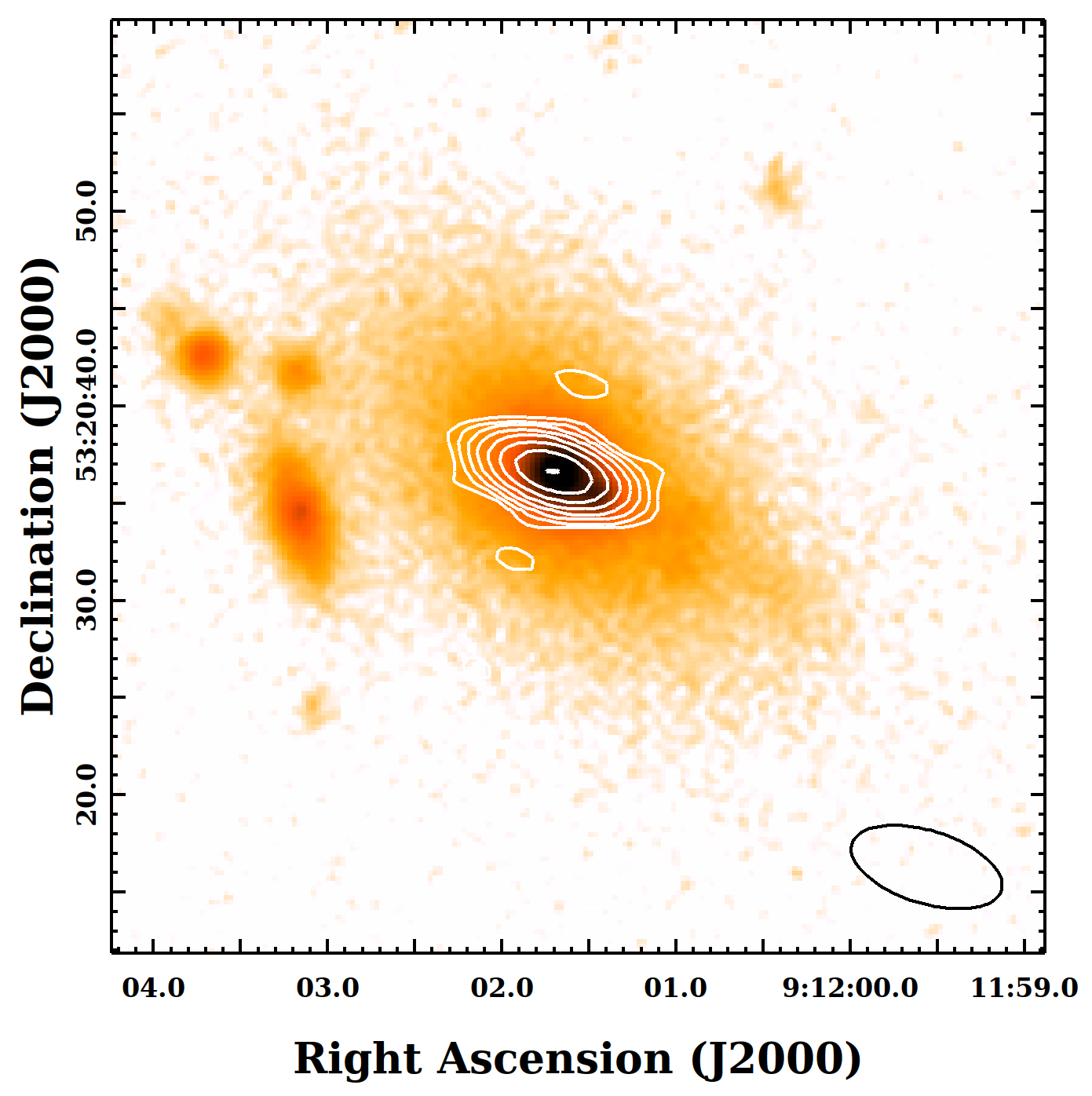}
    \includegraphics[width=0.48\textwidth]{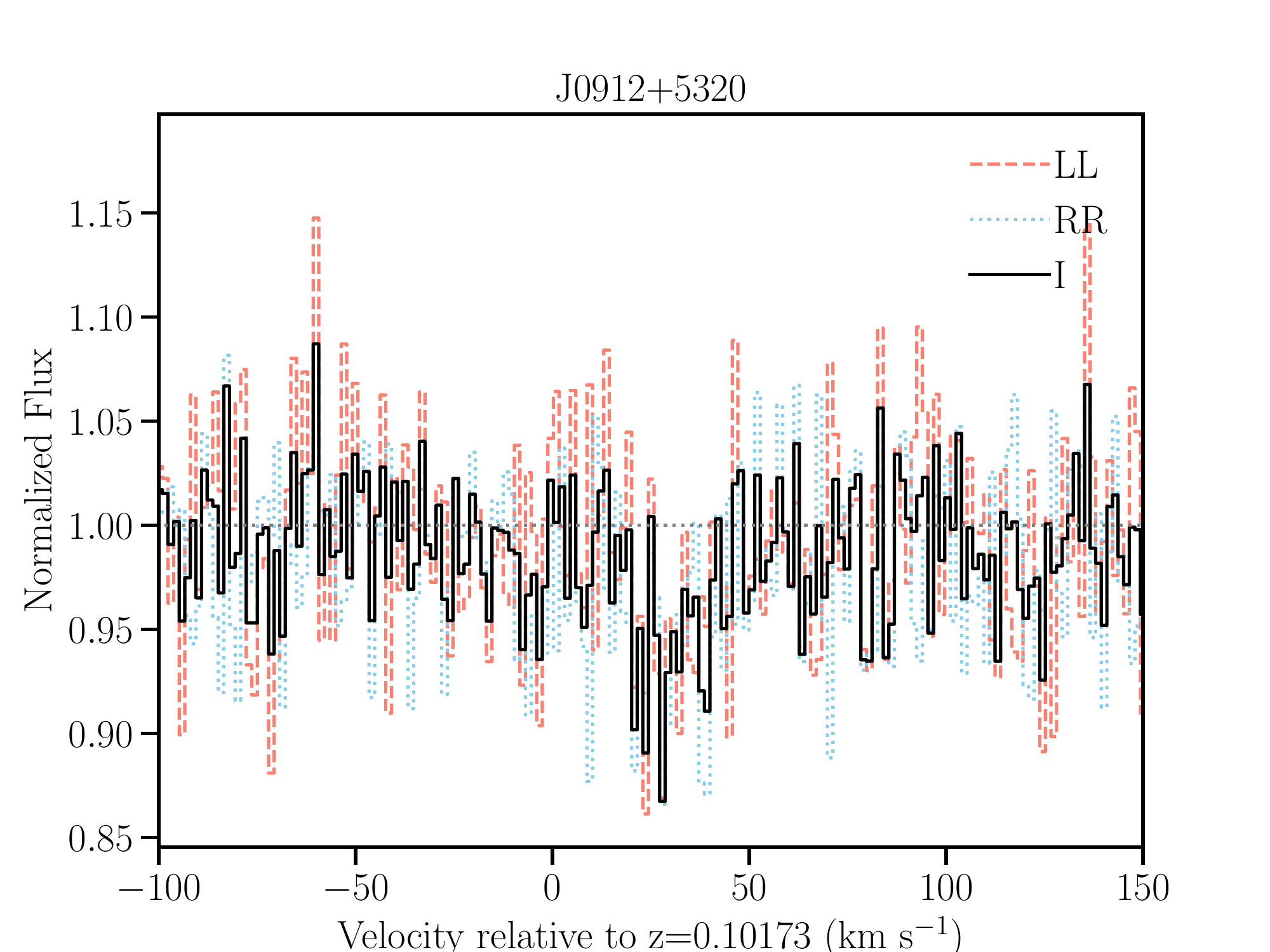}
    \caption{Left: The DECaLS $g$-band images of J0009-0036 (top) and J0912+5320 (bottom). The 1.4\,GHz continuum contours are overlaid in white. The contours are plotted at $C\times$(-1, 1, 2, 4,...)\,\mjb\ where $C$ = 0.3 for J0009-0036 and $C$ = 0.75 for J0912+5320. The restoring beam of the continuum map is shown in the bottom right corner. 
    Right: The tentative \hi\ 21-cm absorption feature detected towards J0009-0036 (top) and J0912+5320 (bottom). The stokes I spectrum is shown in black solid line, while the LL and RR stokes spectra are shown in dashed pink and dotted blue lines, respectively}
    \label{fig:ten_hi_spec}
\end{figure*}

The observations of two of the sources, J0738+3156 and J1352+6541, were affected by strong radio frequency interference (RFI), preventing us from reliably studying the associated \hi\ 21-cm absorption in these cases. The \hi\ 21-cm spectra towards the remaining eight sources, along with the double-peaked \oiii\ emission lines in the SDSS spectra, are shown in Fig.~\ref{fig:sdss_gmrt_spec}. Among the eight sources with reliable \hi\ 21-cm absorption spectra, we detect absorption from three sources (J1356+1026, J1517+3353, J1526+4140). In two cases, J0009-0036 and J0912+5320, we find tentative absorption features at the expected position of the \hi\ 21-cm line. We reobserved the source J0009-0036 using the same observational set-up for an additional 7 hours in October 2019. We combined the spectra of J0009-0036 from the two different observing runs using inverse noise weighting. The peak optical depth of the absorption feature is detected at $3\sigma$ significance in the combined spectrum. In the case of the absorption towards J0912+5320, the peak optical depth is detected at $5\sigma$ significance. However, the absorption lines towards both J0009-0036 and J0912+5320 are narrow (velocity width encompassing 90\% of the total optical depth, \v90\ $\approx30-40$\,\kms), whereas most associated \hi\ 21-cm absorption lines tend to broader on average ($\approx100$\,\kms). Further, the absorption profiles in the stokes LL and RR spectra do not resemble each other perfectly, although they are consistent within the uncertainties. Since there could be low-level sporadic RFI affecting the absorption features, we consider these as tentative detections for our analysis. The two tentative \hi\ 21-cm absorption lines are shown in Fig.~\ref{fig:ten_hi_spec}. For the remaining three sources, J0842+0547, J1203+1319, J1243-0058, where we do not detect absorption, we place $3\sigma$ upper limit on the integrated optical depth for a velocity width of 100\,\kms. Using the integrated optical depth measurement or upper limit, we estimate the corresponding \nhi, assuming a spin temperature, \ts\ = 100\,K, and a covering factor of the radio source, \fc\ = 1. The results from the \hi\ 21-cm observations are provided in Table~\ref{tab_radio_abs}.

\subsection{Incidence of \hi\ 21-cm absorption}
\label{sec_detrate}
Considering the eight sources with clean \hi\ 21-cm absorption spectra and the three confirmed \hi\ 21-cm absorption lines, the detection rate in the sample of double-peaked radio-loud AGN is $38^{+36}_{-20}$\%\footnote{Errors represent Gaussian $1\sigma$ confidence intervals from Poisson statistics following \citet{gehrels1986}.}, which will be $63^{+37}_{-27}$\% if the two tentative detections are true absorption lines. For comparison, in a study of nearby ($z<0.25$) radio galaxies ($S_{\rm 1.4~GHz} > 30$\,mJy), \citet{maccagni2017} detected \hi\ 21-cm absorption in $27\pm5.5$\% of the sources. Extending this to higher redshifts ($0.25<z<0.4$), \citet{murthy2021} found a detection rate of $\approx$19\%, consistent within the errors with the detection rate at lower redshifts. Higher detection rates have been obtained in different classes of radio galaxies, such as compact radio galaxies \citep[$\approx30-60$\%; e.g.][]{gupta2006,chandola2013,maccagni2017}, dust-rich and high excitation radio galaxies \citep[$\approx40$\%; e.g.][]{maccagni2017,chandola2020}, and interacting or merging radio galaxies \citep[$84\pm15$\%;][]{dutta2018,dutta2019}. The incidence of \hi\ 21-cm absorption in double-peaked radio-loud AGN is consistent, within the large uncertainties, with the values reported in the literature for the general population of low-$z$ radio-loud AGN, as well as with the incidence found in merging galaxies.

Among the eight sources with clean \hi\ 21-cm absorption spectra, four are part of systems that show signatures of interactions and tidal disturbances, and four appear to be single and undisturbed based on the available ground-based optical images. We note that the single galaxies may show signatures of interactions in higher spatial resolution images. However, for the sake of the discussion here, we classify the systems based on features identified in the typical spatial resolution images that are available. Two of the interacting systems give rise to \hi\ 21-cm absorption, while two of them show tentative absorption. On the other hand, only one out of the four single systems show \hi\ 21-cm absorption. The tentative higher detection rate ($\gtrsim50$\%) in the interacting systems compared to the single systems ($\approx25$\%) is consistent with the high incidence of \hi\ 21-cm absorption reported in low-$z$ radio-loud galaxy mergers and supports the picture of elevated \hi\ gas in centres of merging systems \citep{dutta2018,dutta2019}. 

\subsection{Detections of \hi\ 21-cm absorption}
\label{sec_detections}

\begin{figure*}
    \centering
    \includegraphics[width=0.31\textwidth]{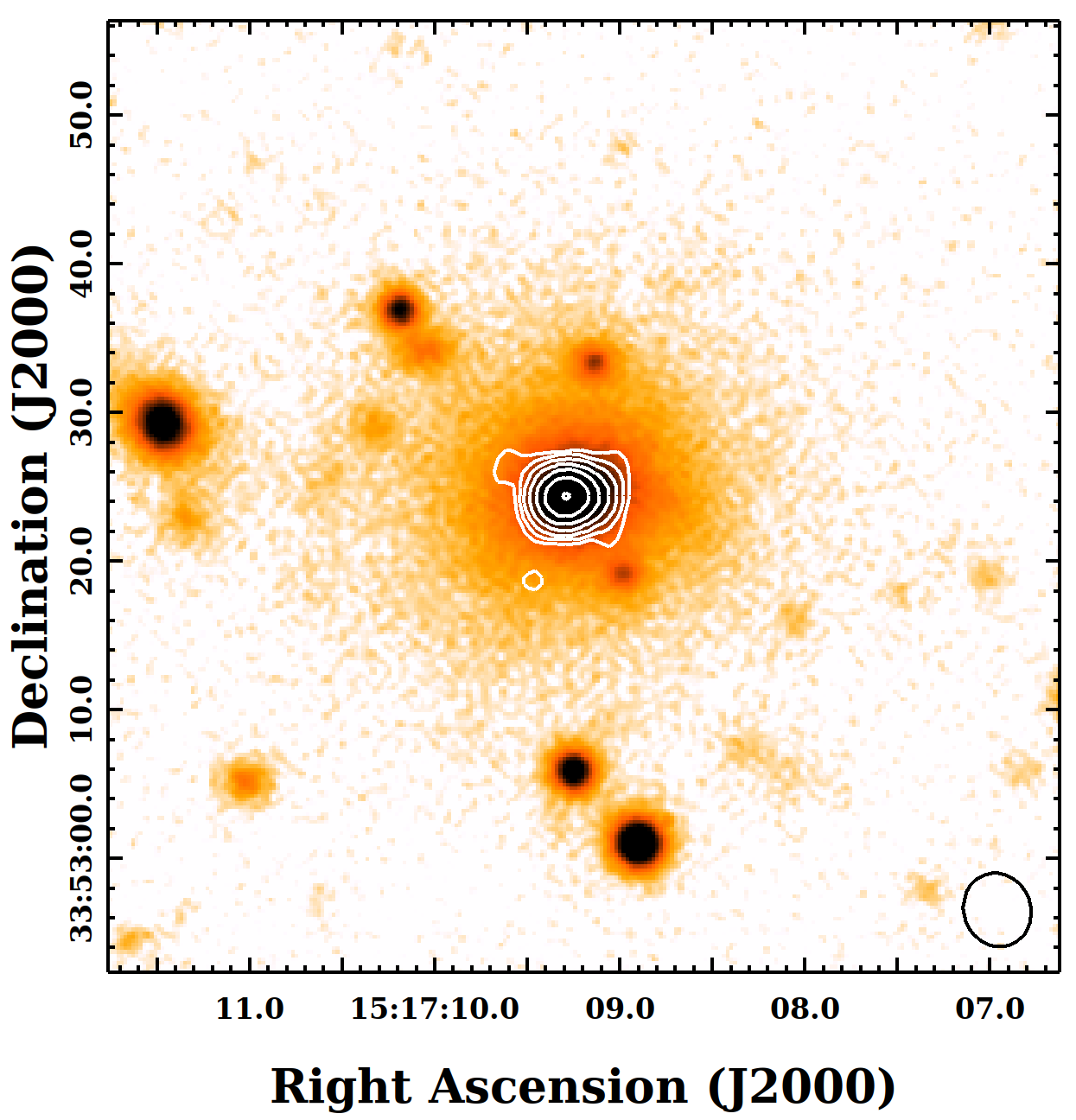}
    \includegraphics[width=0.48\textwidth]{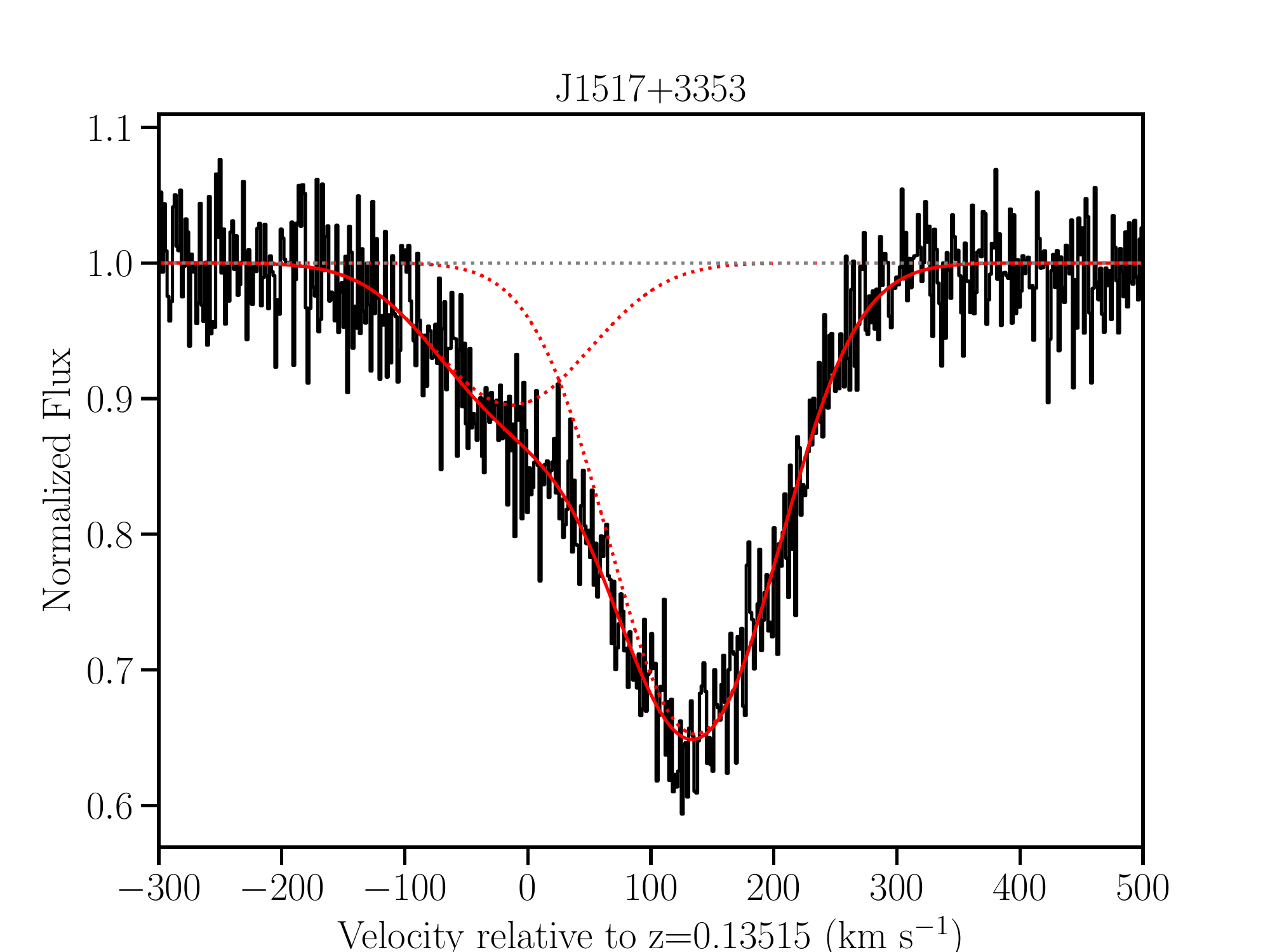}
    \includegraphics[width=0.32\textwidth]{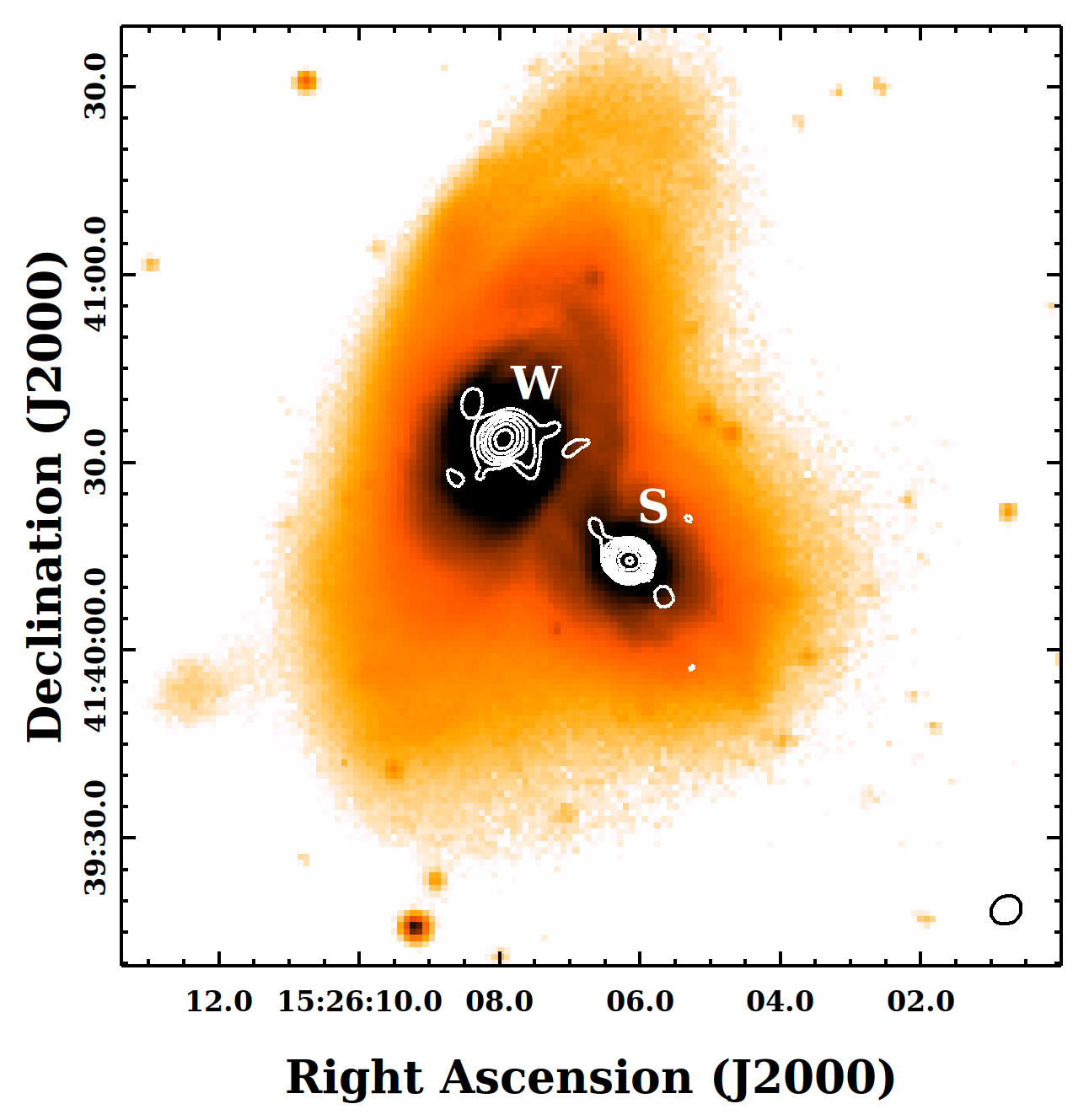}
    \includegraphics[width=0.48\textwidth]{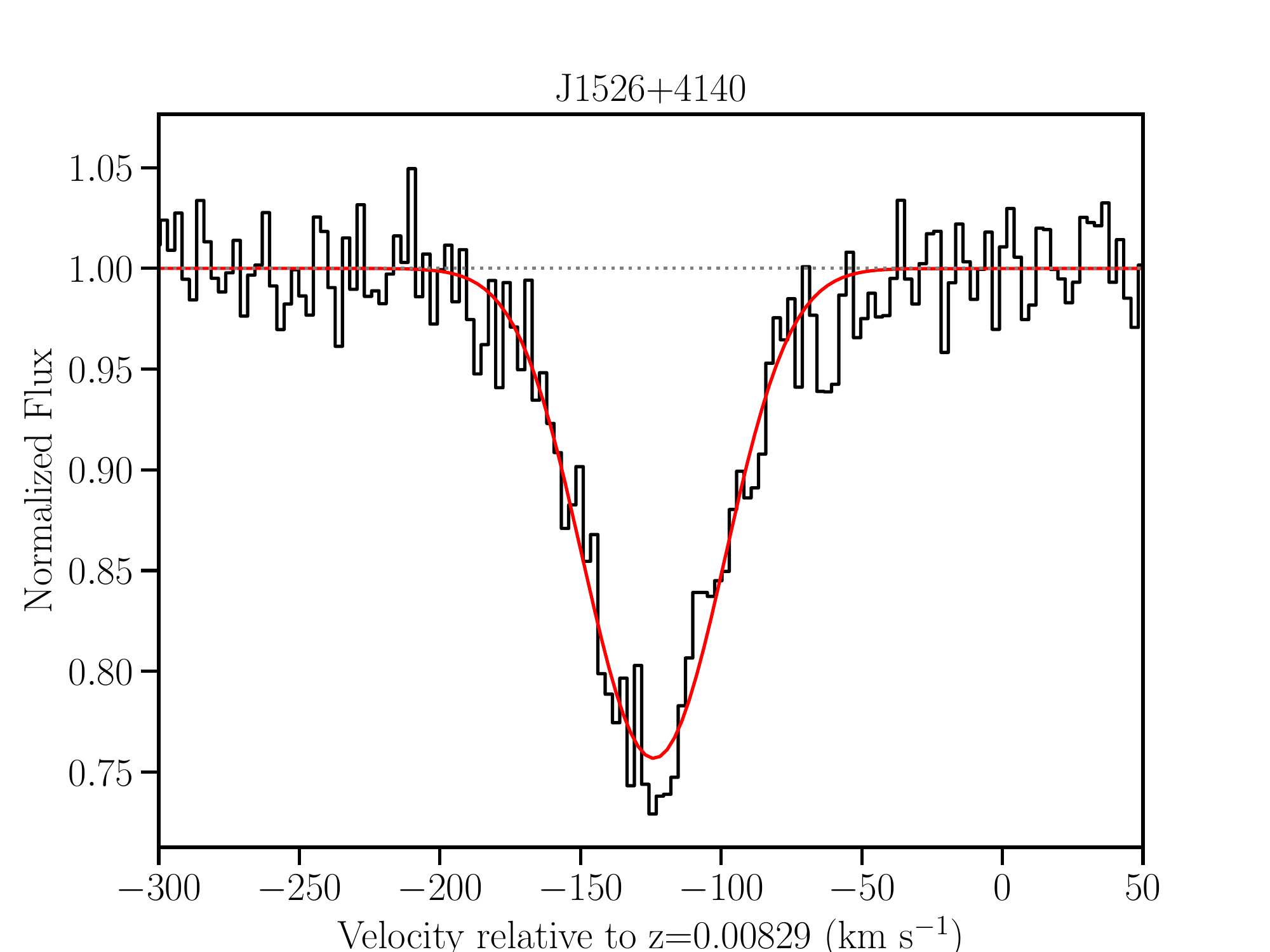}
    \caption{Left: The DECaLS $g$-band images of J1517+3353 (top) and J1526+4140 (bottom). The 1.4\,GHz continuum contours are overlaid in white. The contours are plotted at $C\times$(-1, 1, 2, 4,...)\,\mjb\ where $C$ = 1.5 for J1517+3353 and $C$ = 0.3 for J1526+4140. The restoring beam of the continuum map is shown in the bottom right corner.
    Right: The \hi\ 21-cm absorption detected towards J1517+3353 (top) and J1526+4140 (bottom) in black line. In case of J1526+4140, the \hi\ 21-cm absorption is detected towards the stronger of the two radio sources (marked by `S'). No \hi\ 21-cm absorption is detected towards the weaker radio source (marked by `W'). The best-fitting Gaussian profile is overplotted in red solid line. In case of J1517+3353, the individual Gaussian components are plotted in dotted red line.}
    \label{fig:det_hi_spec}
\end{figure*}

The \hi\ 21-cm absorption detected towards J1356+1026 is discussed in detail in \citet[][see their figure 2 and section 3.1.8]{dutta2018}. Briefly, the absorption has a velocity width encompassing 90\% of the total optical depth (\v90) of 216\,\kms, and corresponds to \nhi\ $\approx5\times10^{21}$\,\cms\ for \ts\ = 100\,K and \fc\ = 1. The \hi\ 21-cm absorption profile is best fit with a single Gaussian component. The location of the peak optical depth is redshifted by $\approx$150\,\kms\ from the systemic redshift that corresponds to average of the two \oiii\ emission lines (see Fig.~\ref{fig:sdss_gmrt_spec}), and is aligned with the stronger of the two \oiii\ emission peaks. We note that the frequency range over which blueshifted \hi\ 21-cm absorption could be detected in this system is affected by RFI.

The \hi\ 21-cm absorption towards J1517+3353 (see top panel of Fig.~\ref{fig:det_hi_spec}) is best fit with two Gaussian components. The strongest component is redshifted by $\approx$136\,\kms\ from the systemic redshift (based on average of the two \oiii\ lines), while the weaker component is at $-13$\,\kms. The absorption arises kinematically between the two \oiii\ emission peaks. The velocity width (\v90) of the absorption is $\approx$281\,\kms. This is one of the strongest \hi\ 21-cm absorption detected in low-$z$ radio-loud AGN, where the integrated optical depth corresponds to \nhi\ $\approx2\times10^{22}$\,\cms\ for \ts\ = 100\,K and \fc\ = 1. Recall that the double-peaked emission in this system is classified as likely originating due to outflows based on optical long-slit spectroscopy \citep{comerford2018}. In addition, multi-band (1.4, 5, 8 and 22 GHz) VLA images show bipolar radio jets aligned with the ionized gas, further supporting the radio jet-driven outflow scenario \citep{rosario2010b}. The weaker, blueshifted \hi\ absorption component could be tracing the neutral outflowing gas in this case.

The \hi\ 21-cm absorption towards J1526+4140 (see bottom panel of Fig.~\ref{fig:det_hi_spec}) is best fit with a single Gaussian component and has a velocity width of $\approx$99\,\kms. The location of the peak optical depth is blueshifted by $\approx$124\,\kms\ from the systemic redshift, and is aligned with the weaker of the two \oiii\ emission peaks. The total optical depth leads to \nhi\ $\approx3\times10^{21}$\,\cms\ for \ts\ = 100\,K and \fc\ = 1. This radio galaxy, also known as NGC~5929, is part of an interacting pair called as Arp~090. Based on high-resolution (25\,pc) MERLIN 1.4\,GHz observations, the radio source consists of a weak core and two lobes \citep{cole1998}. \hi\ 21-cm absorption is detected towards the north-eastern radio lobe with \nhi\ of $6\times10^{21}$\,\cms, while no absorption is detected towards the core (\nhi\ $<3\times10^{21}$\,\cms) or the other lobe (\nhi\ $<0.2\times10^{21}$\,\cms). Assuming that all of the \hi\ 21-cm absorption detected using uGMRT arises from the north-eastern lobe, comparison of the two \nhi\ estimates obtained using uGMRT and MERLIN implies a gas covering factor of $\approx$0.5. \citet{rosario2010a} have reported detection of shocked ionized gas due to radio-jet interaction in this AGN based on Hubble Space Telescope spectra and 5\,GHz radio image. The \hi\ absorption could be tracing either neutral gas outflow, disturbed gas in the circumnuclear region due to the merger occurring in this system, or the large-scale patchy \hi\ gas in the host galaxy that is detected against one of the radio lobes.

\section{Discussion and Summary}
\label{sec_summary}

We have presented a pilot survey of \hi\ 21-cm absorption using uGMRT in $z\le0.4$ radio-loud AGN (median redshift of $\approx0.14$) that exhibit double-peaked \oiii\ emission lines in the SDSS optical spectra. The available ground-based optical images indicate that these AGN are either part of galaxy pairs and interacting systems with tidal tails, or single and undisturbed systems. We were able to search for \hi\ 21-cm absorption in eight double-peaked AGN, and found three detections of \hi\ 21-cm absorption and two tentative detections. As discussed above, the \hi\ 21-cm absorption lines could be tracing neutral gas that is funnelled to the circumnuclear regions during a merger process, feedback from the radio-loud AGN in the form of neutral gas outflows, or neutral gas rotating in circumnuclear or large-scale discs. Higher spatial resolution optical and radio spectroscopy is required to confirm the exact origin of the \hi\ gas detected in absorption.

Based on the confirmed detections, the incidence of \hi\ 21-cm absorption in this sample of double-peaked AGN is $38^{+36}_{-20}$\%. This is consistent within the uncertainties with the typical detection rates of \hi\ 21-cm absorption ($\approx20-30$\%) found in samples of low-$z$ radio-loud AGN in the literature. However, the incidence of \hi\ 21-cm absorption among the systems that show signatures of interaction or tidal disturbance ($\gtrsim50$\%) in the ground-based optical images tends to be higher than that among the single systems ($\approx25$\%). This is in line with the results of the survey of \hi\ 21-cm absorption in radio-loud galaxy mergers at $z\le0.2$ \citep{dutta2018,dutta2019}. This survey found a high incidence of $84\pm15$\% in the mergers that were selected based on visual inspection of SDSS optical images. These results are further consistent with the three times higher atomic gas fractions found in post-mergers relative to a control sample of isolated galaxies in the \hi\ 21-cm emission study of \citet{ellison2018} [however, see also \citet{yu2022} who find that the atomic gas fraction of major-merger pairs on average is marginally decreased by $\approx15$\% relative to isolated galaxies]. The results presented here reinforce the picture that interacting galaxies are likely to be closely linked with the presence of \hi\ gas in their centres.

On going to higher redshifts ($z\gtrsim0.2$), it becomes difficult to directly identify galaxy mergers based on current ground-based images. Upcoming wide-sky surveys such as the Legacy Survey of Space and Time \citep{abell2009} and Euclid \citep{laureijs2011} will provide a much larger sample of galaxy merger candidates out to higher redshifts that can be cross-matched with radio surveys such as the Very Large Array Sky Survey \citep{lacy2020} to identify galaxy mergers hosting radio-loud AGN. Double-peaked emission lines such as \oiii\ in the optical spectra of AGN could in principle provide an alternative technique of identifying dual AGN and galaxy mergers out to $z\approx1$. However, other physical processes such as outflows, ring or discs of star-formation, and complex NLR kinematics could also be responsible for the double-peaked emission. The overall lower incidence of \hi\ gas in the double-peaked AGN sample presented here, compared to the incidence in systems selected based purely on merging activity, also suggests that the presence of double-peaked \oiii\ emission is not necessarily always associated with dual AGN in interacting system. Thus, the results presented here support previous works that find that the double-peaked \oiii\ emission line on its own does not appear to be an efficient selection technique to identify galaxy mergers candidates out to higher redshifts \citep[e.g.][]{shen2011,fu2012,mcgurk2015,comerford2018,liu2018a,rubinur2019}. \hi\ 21-cm absorption observations of a larger sample of double-peaked AGN are required to confirm these results and establish what fraction of double-peaked AGN are likely to be associated with galaxy mergers and nuclear \hi\ gas.

\section*{Acknowledgements}

We thank the anonymous reviewer for helpful comments. We thank the staff at GMRT for their help during the observations. GMRT is run by the National Centre for Radio Astrophysics of the Tata Institute of Fundamental Research. RD gratefully acknowledges support from the European Research Council (ERC) under the European Union’s Horizon 2020 research and innovation programme (grant agreement No 757535), and hospitality from IUCAA where a part of the work was done.

Funding for the Sloan Digital Sky Survey IV has been provided by the Alfred P. Sloan Foundation, the U.S. Department of Energy Office of Science, and the Participating Institutions. SDSS-IV acknowledges support and resources from the Center for High Performance Computing  at the University of Utah. The SDSS website is www.sdss.org. SDSS-IV is managed by the Astrophysical Research Consortium for the Participating Institutions of the SDSS Collaboration including the Brazilian Participation Group, the Carnegie Institution for Science, Carnegie Mellon University, Center for Astrophysics | Harvard \& Smithsonian, the Chilean Participation Group, the French Participation Group, Instituto de Astrof\'isica de Canarias, The Johns Hopkins University, Kavli Institute for the Physics and Mathematics of the Universe (IPMU) / University of Tokyo, the Korean Participation Group, Lawrence Berkeley National Laboratory, Leibniz Institut f\"ur Astrophysik Potsdam (AIP),  Max-Planck-Institut f\"ur Astronomie (MPIA Heidelberg), Max-Planck-Institut f\"ur Astrophysik (MPA Garching), Max-Planck-Institut f\"ur Extraterrestrische Physik (MPE), National Astronomical Observatories of China, New Mexico State University, New York University, University of Notre Dame, Observat\'ario Nacional / MCTI, The Ohio State University, Pennsylvania State University, Shanghai Astronomical Observatory, United Kingdom Participation Group, Universidad Nacional Aut\'onoma de M\'exico, University of Arizona, University of Colorado Boulder, University of Oxford, University of Portsmouth, University of Utah, University of Virginia, University of Washington, University of Wisconsin, Vanderbilt University, and Yale University.

%%%%%%%%%%%%%%%%%%%%%%%%%%%%%%%%%%%%%%%%%%%%%%%%%%

\section*{Data Availability}

The data are available at the GMRT online archive \url{https://naps.ncra.tifr.res.in/goa/data/search}.



\bibliographystyle{mnras}
\bibliography{mybib} % if your bibtex file is called example.bib

%\begin{thebibliography}{99}
%\end{thebibliography}

%%%%%%%%%%%%%%%%%%%%%%%%%%%%%%%%%%%%%%%%%%%%%%%%%%

%%%%%%%%%%%%%%%%% APPENDICES %%%%%%%%%%%%%%%%%%%%%

%\appendix

%%%%%%%%%%%%%%%%%%%%%%%%%%%%%%%%%%%%%%%%%%%%%%%%%%

% Don't change these lines
\bsp	% typesetting comment
\label{lastpage}
\end{document}